\documentclass{appolb}
\usepackage{psfrag,epsfig,graphicx,graphics,xcolor}
\usepackage{wrapfig}
\usepackage{epsfig}

\def\lsim{\raise0.3ex\hbox{$<$\kern-0.75em\raise-1.1ex\hbox{$\sim$}}}
\def\gsim{\raise0.3ex\hbox{$>$\kern-0.75em\raise-1.1ex\hbox{$\sim$}}}

\def\bei{\begin{itemize}}
\def\ei{\end{itemize}}
\def\bea{\begin{eqnarray}}
\def\eea{\end{eqnarray}}
\def\beas{\begin{eqnarray*}}
\def\eeas{\end{eqnarray*}}
\def\beqas{\begin{eqnarray*}}
\def\eqas{\end{eqnarray*}}
\def\beq{\begin{equation}}
\def\eeq{\end{equation}}
\def\beqd{\begin{displaymath}}
\def\eeqd{\end{displaymath}}
\def\eqd{\end{displaymath}}

\def\beeq{\begin{eqnarray}} \def\eeeq{\end{eqnarray}}

\def\bef{\begin{frame}}

\def\slashchar#1{\setbox0=\hbox{$#1$}
   \dimen0=\wd0
   \setbox1=\hbox{/} \dimen1=\wd1
   \ifdim\dimen0>\dimen1
      \rlap{\hbox to \dimen0{\hfil/\hfil}}
      #1
   \else
      \rlap{\hbox to \dimen1{\hfil$#1$\hfil}}
      /
   \fi}

\newcommand{\be}{\begin{equation}}
\newcommand{\ee}{\end{equation}}
\newcommand{\eq}{\end{equation}}
\newcommand{\rb}{\underline{r}}
\newcommand{\kb}{\underline{k}}

\newcommand{\fV}{f_{3\,\rho}^V}
\newcommand{\fA}{f_{3\,\rho}^A}

\newcommand{\zV}{\zeta_{3}^V}
\newcommand{\zA}{\zeta_{3}^A}

\newcommand{\fin}{\end{document}}

\begin{document}
\title{QCD factorization beyond leading twist in exclusive  $\rho_T$-meson production%
\thanks{Talk at Cracow Epiphany Conference on Hadron Interactions at the Dawn of LHC Era. Dedicated to the memory of Jan Kwieci\'nski. 5-7 January 2009 }%
}
\author{I.V.~Anikin$^a$, D.Yu. Ivanov$^b$, B.~Pire$^c$, L.~Szymanowski$^d$, S.~Wallon$^e$
\address{$^a$BLTP, JINR, 141980 Dubna, Russia}
\address{$^b$Institute of Mathematics, 630090 Novosibirsk, Russia}
\address{$^c$CPhT, \'Ecole Polytechnique, CNRS, 91128 Palaiseau, France}
\address{$^d$Soltan Institute for Nuclear Studies, 00-689 Warsaw, Poland} 
\address{$^e$LPT, Universit\'e Paris XI, 91404 Orsay, France}
}

\maketitle
\begin{abstract}
We describe  hard exclusive processes involving a transversally polarized $\rho$ meson in the twist 3
approximation, in a framework based on
the Taylor expansion of the amplitude around the dominant light-cone direction.
\end{abstract}
  
\section{Introduction}

 The last 15 years have witnessed a tremendous   development of the QCD understanding of hard exclusive processes. The basis of this progress has been the derivation of factorization proofs \cite{fact} for exclusive amplitudes in various generalized Bjorken kinematical regimes, at the leading twist level. Schematically, the amplitude of a process governed by a large energy scale $Q$ is written as the convolution of a perturbatively calculable subprocess amplitude and a few hadronic matrix elements of light cone operators. For instance, near forward hard electroproduction of a $\rho$ meson depends on the $H(x,\xi,t) $ and $E(x,\xi,t)$ GPDs and on the $\rho$ distribution amplitudes (DA) $\Phi^\rho(z)$. 
 
 To extend these factorization procedures beyond the leading twist is mandatory in a number of cases including the phenomenologically important instance of the production of transversally polarized vector mesons. 
 The understanding of the quark-gluon structure of a
vector meson is  an important task of hadronic physics if one cares about studying
confinement dynamics. This quark gluon structure may be described by distribution amplitudes which have been discussed in great detail \cite{BB}.

Contrarily to the longitudinally polarized  case, the leading twist 2 DA of the  transversally polarized vector mesons is chirally-odd, a property which opens the way to its use for the study of chirally-odd GPDs \cite{COGPD} (although it is often decoupled because of the vanishing of hard amplitudes \cite{DGP} ).
One thus needs to calculate amplitudes at the twist 3 level  \cite{MP,AT} and this is the source of quite a number of subtle problems, as already discussed in \cite{fact}. Two issues are worth emphasizing. Firstly, new non-perturbative objects enter the discussion, as for instance matrix elements of operators containing a transverse derivative. 
Secondly, the end-point singular behaviour requires careful treatment of kinematical region  with soft partons.     


Both difficulties should be carefully addressed in turn. Here, we shall not discuss the second one, which may hopefully be treated from a modified collinear factorization point of view \cite{BS,GK}, including partonic transverse degrees of freedom and subsequent Sudakov resummation techniques. We plan to return to this question later. The first difficulty has been overviewed by most authors, leading sometimes to the use of gauge dependent expressions for the amplitudes. Indeed preserving gauge invariance in a factorization procedure is not guaranteed, as soon as one allows transverse motion of the partons at some intermediate step of the calculation.  To preserve the QCD gauge invariance, a subtle interplay of various contributing amplitudes must conspire. For instance, in the case we will detail below, QCD gauge invariance is translated to the statement that the  impact factor should vanish when the virtuality of an exchanged
$t-$channel gluon vanishes. We will show that this is indeed the case provided two and three particle correlators are simultaneously taken into account.

Although some courageous attempts have tried to develop the phenomenology of exclusive vector meson production \cite{MP,GK} a careful analysis of the specificities of the transversally polarized case is still lacking.
The existing data from HERA (both at high energy with H1 and ZEUS detectors, and at lower energy with HERMES) on the one side, the forthcoming data from JLab and Compass on the other side, deserve a complete study which we aim to perform within the twist 3 approximation. As a first step \cite{Us}, we show how to use properly the two and three particle distribution amplitudes within a factorized approach.

\section{Calculation of the $\gamma^*\to  \rho_T$ impact factor}
As the simplest example of a hard exclusive reaction involving a transverally polarized  vector meson, let us consider the  high energy  process 
\begin{eqnarray}
\label{prgg}
\gamma^*(q)+\gamma^*(q^\prime)\to \rho_T(p_1)+\rho(p_2)
\end{eqnarray}
where the photons are highly  virtual : $q^2,q'^2=-Q^2,-Q'^2 \gg \Lambda^2_{QCD}$\,, and
the Mandelstam variables  obey the conditions
$s\gg Q^2,\,Q^{\prime\, 2}, -t $. 
Neglecting meson masses, one considers for reaction (\ref{prgg}) the vector meson momenta as  the light cone vectors     ($2\,p_1\cdot p_2=s$);
the "plus" light cone direction being directed along $p_1$ and
 the ``minus'' light cone direction along $p_2$ with vector $n$ 
 defined as $p_2/(p_1 \cdot p_2)$. In this Sudakov  basis, transverse euclidian momenta
are denoted with underlined letters. The virtual photon momentum $q$  reads
$ q=p_1-\frac{Q^2}{2}\, n$.
The impact representation of the scattering amplitude ${\cal M}$ for the reaction (\ref{prgg})  is
\begin{eqnarray}
\label{BFKLamforward}
\frac{i s}{(2\pi)^2}\!\!
\int\frac{d^2\kb}{\kb^2} \Phi^{ab}_1(\kb,\,\rb-\kb) \!\!
\int\frac{d^2\kb'}{\kb'^2} \Phi^{ab}_2(-\kb',\,-\rb+\kb') \!\!\!\!\!
\int\limits_{\delta-i\infty}^{\delta+i\infty} \frac{d\omega}{2\pi i}
\biggl(\frac{s}{s_0}\biggr)^\omega G_\omega (\kb,\kb',\rb)  \nonumber
\end{eqnarray}
where  $G_\omega$ is the 4-gluons Green function which obeys the BFKL equation. $G_\omega$ reduces to $1/\omega \,\delta(\kb -\kb') \kb^2/(\rb-\kb)^2$ within 
Born approximation. 
We focus here on the $\gamma^* \to \rho $ impact factor $\Phi$  
which is the integral of the $\kappa$-channel discontinuity of the  S-matrix element  
 ${\cal S}^{\gamma^*_T\, g\to\rho_T\, g}_\mu$ of the  subprocess $ g(k_1,\varepsilon_{1})+\gamma^*(q)\to g(k_2, \varepsilon_{2})+\rho_T(p_1)$
\begin{eqnarray}
\label{imfac}
\Phi^{\gamma^*\to\rho}(\kb,\,\rb-\kb)= e^{\gamma^*\mu}\, \frac{1}{2s}\int\frac{d\kappa}{2\pi}
\, \hbox{Disc}_\kappa \,  {\cal S}^{\gamma^*\, g\to\rho\, g}_\mu(\kb,\,\rb-\kb)\,,
\end{eqnarray}
where $\kappa=(q+k_1)^2$. Considering  the forward limit for simplicity, 
the gluon momenta read
\begin{eqnarray}
\label{gmom}
k_1=\frac{\kappa+Q^2+\kb^2}{s} p_2 + k_T, \quad
k_2=\frac{\kappa+\kb^2}{s} p_2 + k_T, \quad
k_1^2=k_2^2=k^2_T=-\kb^2\,. \nonumber
\end{eqnarray}

The impact factor $\Phi$  can be calculated within the
collinear factorization. It is a convolution of
 perturbatively calculable hard-scattering amplitudes $H$ and soft correlators $S$ involving relevant $\rho$-meson
distribution amplitudes, symbolically written as
\begin{equation}
\label{factorisation}
\Phi= \int d^4l \cdots \, \hbox{tr} [H(l \cdots) \, S(l \cdots)]\,,
\end{equation}
and which involves loop integrations formed by the $n$ partons ($n \ge 2$) 
entering soft correlators $S$.
 Working within the  light-cone collinear factorization framework, let us first derive the contribution of the diagrams with the quark-antiquark
correlators. The basic tool is to decompose any hard  coefficient function $H(l)$  around
a dominant ``plus" direction:
\begin{equation}
\label{expand}
H(\ell) = H(y p) + \frac{\partial H(\ell)}{\partial \ell_\alpha} \biggl|_{\ell=y p}\biggr. \,
(\ell-y\, p)_\alpha + \ldots\,,
\end{equation}
with $(\ell-y \, p)_\alpha \approx \ell^\perp_\alpha $ up to twist 3 accuracy. Here and subsequently $\alpha$ denotes transverse components. This collinear expansion of the hard part should be compared to the expansion of matrix elements in powers of space-time separation $(x^2)^n$ in the
coordinate approach.

\begin{figure}[t]
\psfrag{l}[cc][cc]{$\ell$}
\psfrag{lm}[cc][cc]{}
\psfrag{q}[cc][cc]{$\gamma^*$}
\psfrag{H}[cc][cc]{$H$}
\psfrag{S}[cc][cc]{$\Phi$}
\psfrag{Hg}[cc][cc]{$H_\mu$}
\psfrag{Sg}[cc][cc]{$\Phi^\mu$}
\psfrag{k}[cc][cc]{}
\psfrag{rmk}[cc][cc]{}
\psfrag{rho}[cc][cc]{$\rho$}
\centering{
\begin{tabular}{cccc}
\includegraphics[width=3.8cm]{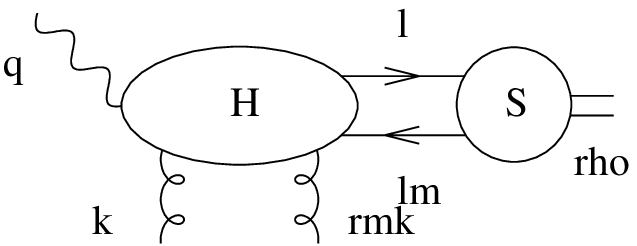}&\hspace{-.2cm}
\raisebox{.7cm}{+}&\hspace{-.1cm}\includegraphics[width=3.8cm]{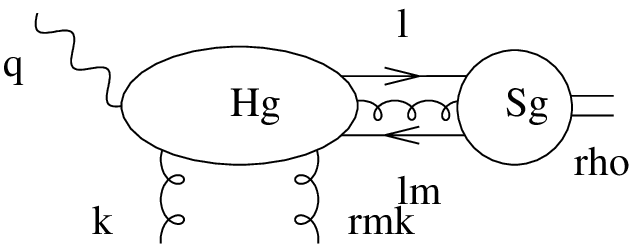}
&\hspace{-.2cm}\raisebox{.7cm}{$+ \cdots$}
\end{tabular}
}
\caption{2- and 3-body correlators attached to a hard scattering amplitude in the  example of the $\gamma^* \to \rho$ impact factor.}
\label{fig:NonFactorized}
\end{figure}

The $l^\perp$ dependence of the hard part looks first as an excursion out of
the collinear framework.
The factorized expression is obtained as the result of an integration by parts
which replaces  $\ell^\perp_\alpha$ by $\partial^\perp_\alpha$ acting on
the soft correlator.
 This leads to new operators
${\cal O}^\perp$ which contain
transverse derivatives, such as $\bar \psi \, \partial^\perp \psi $,
and thus
to the necessity of considering additional correlators
$\Phi^\perp (l)$.
This procedure leads to the factorization of the amplitude in momentum
space. Factorization in the Dirac space can be achieved by
a Fierz decomposition. 
Thus,  the amplitude  takes the  factorized form, see Fig.~1:

\begin{eqnarray}
\label{GenAmpFac}
{\cal A}=
\int\limits_{0}^{1} dy \,{\rm tr} \left[ H(y) \, \Gamma \right] \,
\Phi^{\Gamma} (y)\,
+\int\limits_{0}^{1} dy_1\, dy_2 \,{\rm tr} \left[ H^\mu(y_1,y_2)
\, \Gamma \right] \, \Phi^{\Gamma}_{\mu} (y_1,y_2) .
\end{eqnarray}
The parameterizations of
the vacuum--to--$\rho$-meson matrix elements needed
in the above factorization procedure up to the twist-$3$ order
in the axial (light-like) gauge $n \cdot A=0$,
which we will use in order to get rid of Wilson line,
can be written
as (here $z=\lambda n$) 
\begin{eqnarray}
\label{CorrelatorV}
\langle \rho(p)|\bar\psi(z)\gamma_{\mu} \psi(0)|0\rangle
& \stackrel{{\cal F}_1}{=}& m_\rho \, f_\rho \,[
\varphi_1(y)\, (e^*\cdot n)p_{\mu}+\varphi_3(y)\, e^{*\,T}_{\mu}]\,, \nonumber
\\
\label{CorrelatorA}
\langle \rho(p)|
\bar\psi(z)\gamma_5\gamma_{\mu} \psi(0) |0\rangle
&\stackrel{{\cal F}_1}{=}&m_\rho \, f_\rho \,
i\varphi_A(y)\,
\varepsilon_{\mu\lambda\beta\delta}\,
e^{*\,T}_{\lambda}\, p_{\beta} \, n_{\delta}
\end{eqnarray}
and
\begin{eqnarray}
\label{CorrelatorDerV}
\hspace{-.5cm}\langle \rho(p)|
\bar\psi(z)\gamma_{\mu}
i\stackrel{\longleftrightarrow}
{\partial^\perp_{\alpha}} \psi(0)|0 \rangle
&\stackrel{{\cal F}_1}{=}&m_\rho \, f_\rho \,
\varphi_1^T(y) \, p_{\mu} e^{*\,T}_{\alpha} \, ,\nonumber
\\
\label{CorrelatorDerA}
\langle \rho(p)| \bar\psi(z)\gamma_5\gamma_{\mu}
i\stackrel{\longleftrightarrow}
{\partial^\perp_{\alpha}} \psi(0) |0\rangle
&\stackrel{{\cal F}_1}{=}&m_\rho \, f_\rho \,
i \, \varphi_A^T (y) \, p_{\mu} \,
\varepsilon_{\alpha\lambda\beta\delta}
\, e^{*\,T}_{\lambda} \, p_{\beta} \, n_{\delta}\,,
\end{eqnarray}
where
$\stackrel{{\cal F}_1}{=}$
denotes the Fourier transformation
$\int_{0}^{1}\, dy e^{\left[iy\,p\cdot z\right]}\,.$
The matrix elements of quark-gluon nonlocal operators can be
parameterized as
\begin{eqnarray}
\label{Correlator3BodyV}
\langle \rho(p)|
\bar\psi(z_1)\gamma_{\mu}g A_{\alpha}^T(z_2) \psi(0) |0\rangle
&\stackrel{{\cal F}_2}{=}&
m_\rho \,\fV\,
B(y_1,y_2)\, p_{\mu} e^{*T}_{\alpha}\,,
\\
\label{Correlator3BodyA}
\langle \rho(p)|
\bar\psi(z_1)\gamma_5\gamma_{\mu} g A_{\alpha}^T(z_2) \psi(0) |0\rangle
&\stackrel{{\cal F}_2}{=}&
m_\rho \,\fA\,
i \,
D(y_1,y_2)\, p_{\mu}
\varepsilon_{\alpha\lambda\beta\delta}
\, e^{*T}_{\lambda} \, p_{\beta} \, n_{\delta}\,,\nonumber
\end{eqnarray}
where
$\int\limits_{0}^{1} dy_1 \,\int\limits_{0}^{1} dy_2 \,
e^{\left[ iy_1\,p\cdot z_1+i(y_2-y_1)\,p\cdot z_2 \right]}$
is denoted by $\stackrel{{\cal F}_2}{=}\,.$
The light-cone fractions of the quark,
anti-quark and gluon are  $y_1$, $1-y_2$ and $y_2-y_1\,.$

The correlators introduced above are not independent,
thanks to the QCD equations of motion for the field
operators entering them (see, for example, \cite{AT}).
In the simplest case of fermionic fields, they follow from the vanishing matrix elements
$\langle i  
{\hat D}(0) \psi(0)\, \bar \psi(z)\rangle = 0$ and
$\langle  \psi(0)\, i {\hat D}(z)\bar \psi(z)\rangle
= 0\,$
due to the Dirac equation.
Denoting $\zeta_{3, \, \rho}^{V,A}=f_{3 \,\rho}^{V,A}/f_\rho$, we obtain
\begin{eqnarray}
\label{em_rho1}
 &&\bar{y}_1 \, \varphi_3(y_1) +  \bar{y}_1 \, \varphi_A(y_1)  +  \varphi_1^T(y_1)  +\varphi_A^T(y_1)=\nonumber \\
&&-\int\limits_{0}^{1} dy_2 \left[ \zV \, B(y_1,\, y_2) +\zA \, D(y_1,\, y_2) \right] \,
\end{eqnarray}
and
\begin{eqnarray}
\label{em_rho2}
 && y_1 \, \varphi_3(y_1) -  y_1 \, \varphi_A(y_1)  -  \varphi_1^T(y_1)  +\varphi_A^T(y_1) \nonumber\\
&&=-\int\limits_{0}^{1} dy_2 \left[ -\zV \, B(y_2,\, y_1) +\zA\, D(y_2,\, y_1) \right] \,.
\end{eqnarray}

\begin{figure}[tb]
\psfrag{yq}{{\tiny $\hspace{-.2cm}y_1$}}
\psfrag{dy}{\raisebox{.04cm}{\tiny $\hspace{-.3cm}y_2-y_1$}}
\psfrag{yb}{{\tiny $\hspace{-.2cm}1-y_2$}}
\centering{
\hspace{-0.2cm}\raisebox{1cm}{$(y_1-y_2)$}\  \includegraphics[width=2.5cm]{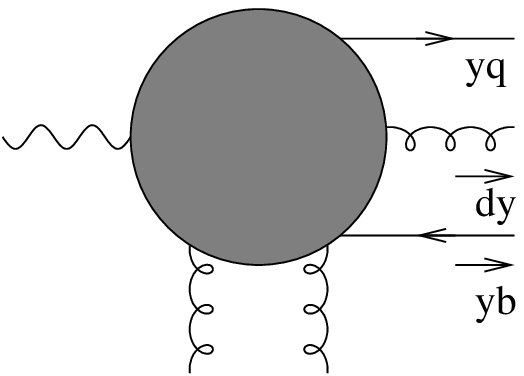}\hspace{.4cm} \raisebox{1cm}{=}\,
\psfrag{yq}{{\tiny $\hspace{-.2cm}y_1$}}
\psfrag{yb}{{\tiny $\hspace{-.2cm}1-y_1$}}
\hspace{0.2cm}\includegraphics[width=2.5cm]{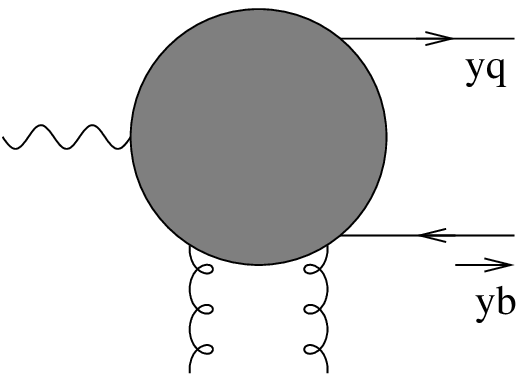}\hspace{0.4cm}\raisebox{1cm}{-}\hspace{.2cm}
\psfrag{yq}{{\tiny $\hspace{-.2cm}y_2$}}
\psfrag{yb}{{\tiny $\hspace{-.3cm}1-y_2$}}
\includegraphics[width=2.5cm]{RWardIF.eps}
}
\caption{Reduction of
 3-body correlators to 2-body correlators through Ward identity.}
\label{fig:WardIF}
\end{figure}

A crucial point of this approach is that
scattering amplitudes do not depend on the specific choice of
the vector $n$ which fixes the light-cone direction.
As observed   for inclusive structure functions \cite{EFP} and for DVCS \cite{KivelRad},
this $n$ independence condition leads
to non-trivial relations at the twist three level.
In the case of exclusive processes \cite{AT}, this constrains the non-perturbative correlators entering the factorized amplitude.
The polarization vector for transverse  $\rho$ which enters in the
parametrization of twist 3 correlators depends on $n$
according to
\begin{equation}
\label{epsilonRho}
e^{*T}_\mu = e^*_\mu - p_\mu\, e^*\cdot n
\end{equation}
i.e. it is defined with respect to the light-cone vector $n$.
The  $n-$independence condition can thus be written \nolinebreak as
\begin{equation}
\label{nInd}
\frac{d{\cal A}}{dn^\mu}
=0\,, \qquad \mbox{where} \quad  \frac{d}{dn^\mu}= \frac{\partial}{\partial n^\mu} + e^*_\mu\frac{\partial}{\partial (e^* \cdot n)}\,.
\end{equation}
The appearance of the total  derivative may be interpreted
as a (vector) analog to the renormalization group (RG) invariance when the
dependences on the renormalization parameter coming from various sources
cancel.  One can view this as a RG-like flow in the space of light-cone
directions of contributions to the amplitude where the polarization vector plays the
role of a beta function.
This  condition expressed at the level of the {\em full amplitude} can be reduced to a set of conditions involving only the soft correlators. 
The general strategy relies on the use of the collinear Ward identities which relate firstly amplitudes with different
number of legs and secondly higher order coefficients in the Taylor expansion (\ref{expand}) to lower order ones.
The most involved use of these identities occurs in the case of
the 3-body correlator.
In the case of vector correlator (\ref{Correlator3BodyV}), due to (\ref{epsilonRho}) the dependency on $n$ enters linearly and only through the scalar product $e^* \cdot n\,.$
 Thus, the action on the amplitude of the derivative $d/dn$ defined in (\ref{nInd}) can be extracted by the replacement $e_\alpha^* \to p_\alpha\,,$ and after using the Ward identity, it reads
\begin{eqnarray}
\label{Ward}
(y_1-y_2){\rm tr} \left[ H_{\rho}(y_1,y_2) \, p^\rho
\, \slashchar{p}\right] =
{\rm tr} \left[H(y_1) \, \slashchar{p}\right]-{\rm tr} \left[H(y_2) \, \slashchar{p}\right]\,,
\nonumber
\end{eqnarray}
as illustrated by
Fig.\ref{fig:WardIF}.
One can apply similar tricks to the 2-body correlators with transverse derivative
whose contribution can be viewed as 3-body processes with vanishing gluon  momentum.
This finally leads  to the equations
\begin{eqnarray}
\label{ninV}
&&\frac{d}{dy_1}\varphi_1^T(y_1)=-\varphi_1(y_1)+\varphi_3(y_1)-\zV\int\limits_0^1\,\frac{dy_2}{y_2-y_1}
 \left( B(y_1,y_2)+B(y_2,y_1) \right)\,, \nonumber \\
&&\frac{d}{dy_1}\varphi_A^T(y_1)=\varphi_A(y_1)-\zA\int\limits_0^1\,\frac{dy_2}{y_2-y_1}
 \left(D(y_1,y_2)+D(y_2,y_1)
   \right) \,.
\end{eqnarray}
We emphasize that the equations obtained above for the special case of impact factor are universal i.e. do
not refer to any specific hard process.
The equations of motion  together with those expressing the
 $n$-independence are four equations which constrain
seven DAs ($\varphi_1,\varphi_3,\varphi_1^T,\varphi_A,\varphi_A^T,B,D$).
Thus, any hard process involving the
exclusive production of a $\rho$ can be expressed, at twist 3, in terms
of the three independent DAs which  we choose as
$\varphi_1$, $B$ and $D$.
The solution of Eqs. (\ref{em_rho1},\,\ref{em_rho2},\,\ref{ninV}) can be generically written as the sum of 
the solution in the Wandzura-Wilczek approximation, i.e. with vanishing $B$ and $D$ DAs, and the solution
describing the genuine twist 3 part.
The Wandzura-Wilczek parts 
take the forms 
\begin{eqnarray}
\label{WWphi3}
\!\varphi_{3/A}^{WW}(y)&\!\!=\!\!&\frac{1}{2}\left[ \int\limits_0^y\,\frac{dv}{\bar v}\varphi_1(v)
\pm\int\limits_y^1\frac{dv}{v}\varphi_1(v)   \right]\,,\\
\label{WWphi1T}
\!\varphi_{1/A}^{T\,WW}(y)&\!\!=\!\!&\frac{1}{2}\left[
-\bar y \! \int\limits_0^y\,\frac{dv}{\bar v}\varphi_1(v)
\pm y\!\int\limits_y^1\frac{dv}{v}\varphi_1(v)   \right]\!.
\end{eqnarray}
The remaining genuine twist 3 DAs read:
\begin{eqnarray}
\label{Resphi3}
&&\varphi_3^{gen}(y)=-\frac{1}{2}\int\limits^1_{y} \frac{du}{u}
\biggl[
\int\limits^u_0 dy_2 \frac{d}{du} (\zV B-\zA D)(y_2,\, u) 
\biggr.
\nonumber\\
\biggl.
&&-\int\limits^1_u \frac{dy_2}{y_2-u}  (\zV B-\zA D)(u, \,y_2) 
-\int\limits^u_0 \frac{dy_2}{y_2-u}  (\zV B-\zA D)(y_2, \, u)
\biggr]
\nonumber \\
&&-\frac{1}{2}\int\limits^{y_1}_{0}  \frac{du}{\bar{u}}
\biggl[
\int\limits^1_u dy_2 \frac{d}{du} (\zV B+\zA D)(u, \,y_2)
\biggr.
\\
\biggl.
&&-\int\limits^1_u \frac{dy_2}{y_2-u}  (\zV B+\zA D)(u, \,y_2) 
-\int\limits^u_0 \frac{dy_2}{y_2-u}  (\zV B+\zA D)(y_2, \, u)
\biggr]\,, \nonumber
\end{eqnarray}
\begin{equation}
\label{Resphi1T}
\varphi_1^{T \, gen}(y) =\int\limits^y_0 du \, \varphi_3^{gen}(u) - \zV \int\limits^y_0 dy_1 \int\limits^1_y dy_2 \frac{B(y_1,\, y_2)}{y_2-y_1}\,,
\end{equation}
while the corresponding expressions for $\varphi^{gen}_A(y)$ and
   $\varphi_A^{T\, gen}(y)$ are obtained by the substitutions:
\begin{eqnarray}
\label{ResphiA}
\varphi_{A}^{gen}(y) &\stackrel{\zV B \,\leftrightarrow \,\zA D}{\longleftrightarrow}& \varphi_{3}^{gen}(y)\, ,\\
\label{ResphiAT}
\varphi_{A}^{T\, gen}(y) &\stackrel{\zV B \,\leftrightarrow \,\zA D}{\longleftrightarrow}& \varphi_{1}^{T\, gen}(y)\,.
\end{eqnarray}

In summary, we describe consistently exclusive
processes at higher twist, in a way which explicitly preserves
gauge invariance. 
The $n$-independency condition generalized up to the dynamical
twist $3$ plays a crucial role for 
the consistency of this approach with the studies of DAs performed in 
 \cite{BB}.
The present framework opens the way to a systematic and consistent treatment of
hard exclusive  processes.
This does not preclude the solution of the 
end-point singularity problem \cite{MP, GK} which  requires a dedicated treatment.


\noindent
We thank O.~V.~Teryaev for discussions.
This work is supported by the Polish Grant N202 249235, the French-Polish scientific agreement Polonium, the grant ANR-06-JCJC-0084, 
the RFBR (grant 09-02-01149 and 07-02-91557, 08-02-00334, 08-02-00896).

\end{document}